\documentclass[%
 aip,
 jmp,%
 amsmath,amssymb,
 reprint,%
]{revtex4-1}
\usepackage{color}
\usepackage{graphicx}% Include figure files
\usepackage{dcolumn}% Align table columns on decimal point
\usepackage{bm}% bold math
\usepackage{psfrag}
\usepackage[normalem]{ulem}
\usepackage{fancyhdr}
\begin{document}
\pagestyle{fancy} \lhead{The paper has been published at J. Appl. Phys. 119, 064506 (2016)} 
%\preprint{AIP/123-QED}

\title[]{Artificial Perfect Electric Conductor-Perfect Magnetic Conductor Anisotropic Metasurface for Generating Orbital Angular Momentum of Microwave with Nearly Perfect Conversion Efficiency}
%\thanks{Footnote to title of article.}

\author{Menglin L.N. Chen}
\affiliation{Department of Electrical and Electronic Engineering, The University of Hong Kong, Pokfulam Road, 00852, Hong Kong.}%

\author{Li Jun Jiang}
\email{ljiang@eee.hku.hk}
\affiliation{Department of Electrical and Electronic Engineering, The University of Hong Kong, Pokfulam Road, 00852, Hong Kong.}%Lines break automatically

 \author{Wei E.I. Sha}
 \email{wsha@eee.hku.hk}
\affiliation{Department of Electrical and Electronic Engineering, The University of Hong Kong, Pokfulam Road, 00852, Hong Kong.}%

\date{\today}% It is always \today, today,

\begin{abstract}
Orbital angular momentum (OAM) is a promising degree of freedom for fundamental studies in electromagnetics and quantum mechanics. The unlimited state space of OAM shows a great potential to enhance channel capacities of classical and quantum communications. By exploring the Pancharatnam-Berry phase concept and engineering anisotropic scatterers in a metasurface with spatially varying orientations, a plane wave with zero OAM can be converted to a vortex beam carrying nonzero OAM. In this paper, we proposed two types of novel PEC (perfect electric conductor)-PMC (perfect magnetic conductor) anisotropic metasurfaces. One is composed of azimuthally continuous loops and the other is constructed by azimuthally discontinuous dipole scatterers. Both types of metasurfaces are mounted on a mushroom-type high impedance surface. Compared to previous metasurface designs for generating OAM, the proposed ones achieve nearly perfect conversion efficiency. In view of the eliminated vertical component of electric field, the continuous metasurface shows very smooth phase pattern at the near-field region, which cannot be achieved by convectional metasurfaces composed of discrete scatterers. On the other hand, the metasurface with discrete dipole scatterers shows a great flexibility to generate OAM with arbitrary topological charges. Our work is fundamentally and practically important to high-performance OAM generation.
\end{abstract}

\maketitle

\section{Introduction}
Electromagnetic momentum density can be decomposed in terms of orbital momentum and spin momentum densities~\cite{bliokh2013dual}. They are respectively responsible for the generation of the orbital angular momentum (OAM) and spin angular momentum (SAM) of electromagnetic (EM) waves. Left and right circularly polarized EM waves carry SAM of $\pm\hbar$ that is intrinsic (origin-independent) physical quantity. Fundamentally different from SAM, OAM is an extrinsic origin-dependent quantity, which can be carried by vortex beams with a helical wavefront~\cite{allen}. On the other hand, the unbounded eigenstates of OAM could enhance capacities of radio, optical and quantum communications~\cite{zhang,radio,wang2012terabit,gibson2004free,mair2001entanglement}. Additionally, OAM has various potential applications involving super-resolution imaging~\cite{bessel}, optical tweezers~\cite{tweezer}, etc.

There are several approaches to generate OAM of EM waves. One common approach is to introduce desired phase retardation by spiral phase plates~\cite{SSP_SREP}, antenna arrays~\cite{mohammadi2010orbital}, holographic plates~\cite{genevet2012holographic}, etc. Another way is to harness abrupt phase changes by exploiting Pancharatnam-Berry phase concept~\cite{APL_spin_to_orbital, LSA_spin_to_orbital, CapassoAPL, reviewer1_CapassoScience, Capasso_review, kang2012twisted, kangming_eot}. Using anisotropic scatterers in a metasurface, with spatially varying orientations, a vortex beam with the helical phase can be created. The main pitfall in current OAM designs by metasurface is the low conversion efficiency from a plane wave with zero OAM to the vortex beam with nonzero OAM. For example, a metasurface composed of V-shaped scatterers with varied geometric parameters~\cite{CapassoAPL, reviewer1_CapassoScience} was proposed to generate OAM in the cross-polarized component of scattered field under a linearly polarized wave incidence. The design achieved a polarization conversion of 30\%. Another example is to employ the aperture antennas~\cite{kangming_eot} that act as linear polarizers. An azimuthally polarized OAM beam was generated under a circularly polarized incident wave. The conversion efficiency limit is bounded by $50\%$.

In this paper, we propose two types of novel PEC (perfect electric conductor)-PMC (perfect magnetic conductor) anisotropic metasurfaces to overcome the low efficiency issue existing in current OAM designs. One of proposed metasurface could perfectly convert a left (right) circularly polarized plane wave carrying zero OAM to a right (left) circularly polarized vortex beam carrying arbitrary order OAM. With azimuthally continuous loops, the other proposed metasurface generates much smoother near-field phase pattern than conventional metasurfaces with discrete scatterers.

\begin{figure*}[!tbc]
\centering
\includegraphics[width=4.2in]{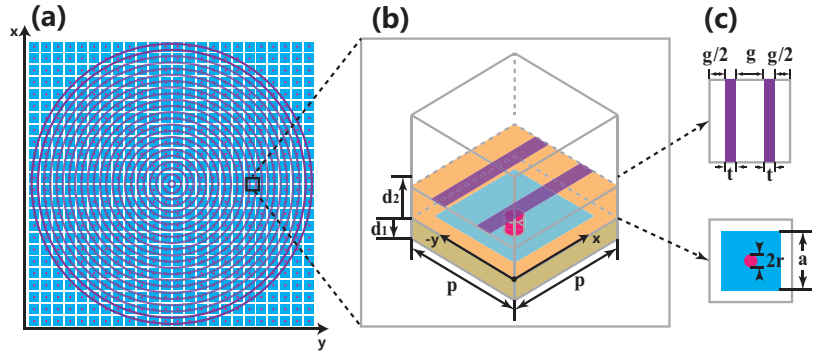}
\caption{Schematic pattern of the PEC-PMC anisotropic metasurface for OAM generation. With a nearly 100\% conversion efficiency, the metasurface perfectly converts a left (right) circularly polarized plane wave carrying zero OAM to a right (left) circularly polarized vortex beam carrying $\pm 2\hbar$ OAM. (a) top view of the whole metasurface; (b, c) a scatterer in the metasurface. The scatterer is composed of artificial PEC (purple) and PMC (blue and red) surfaces. The period of the scatterer is $p=7$ mm. The permittivity and thickness of the dielectric substrate are set to $\epsilon_{r}=2.2$, $d_1=2$~mm and $d_2=3$~mm. For the artificial PEC surface (top-right inset), the width and gap for the strip is $t=1$~mm and $g=2.5$~mm, respectively. For the mushroom based artificial PMC surface (bottom-right inset), the square patch size is $a=6$~mm. A metallic via with the radius of $r=0.25$~mm and height of $d_1=2$~mm connects the patch to the ground plane.}
\label{Fig:unit_cell}
\end{figure*}

\section{Theory}

For an anisotropic scatterer in metasurface, linear polarized Jones vectors of the incident and scattered (transmitted or reflected) fields can be connected by the Jones matrix $\mathbf{\overline{J}}$

\begin{equation}
\begin{pmatrix}
  E^s_x \\ E^s_y
\end{pmatrix}
=
\begin{pmatrix}
    J_{xx} & J_{xy} \\
    J_{yx} & J_{yy}
  \end{pmatrix}
\begin{pmatrix}
  E^i_x \\ E^i_y
  \end{pmatrix}
  =
    \mathbf{\overline{J}}
\begin{pmatrix}
  E^i_x \\ E^i_y
\end{pmatrix}
\label{Eq:1}
\end{equation}
where $E^i_x$ and $E^i_y$ are the $x$ and $y$ components of the incident electric field. $E^s_x$ and $E^s_y$ are the corresponding components of the scattered electric field. If $J_{yy}=-J_{xx}$ and $J_{yx}=J_{xy}$, azimuthally rotating the scatterer by an angle of $\alpha$ will result in a new Jones matrix
\begin{equation}
\begin{split}
&\mathbf{\overline{J}}(\alpha)=\\
&\begin{pmatrix}
  J_{xx}\cos(2\alpha)-J_{xy}\sin(2\alpha) & J_{xx}\sin(2\alpha)+J_{xy}\cos(2\alpha) \\
  J_{xx}\sin(2\alpha)+J_{xy}\cos(2\alpha) & J_{xy}\sin(2\alpha)-J_{xx}\cos(2\alpha)
\end{pmatrix}
\end{split}
\label{Eq:2}
\end{equation}

\noindent Under circular basis, $\mathbf{\overline{J}}(\alpha)$ will convert to
\begin{equation}
\mathbf{\overline{J}}_{c}(\alpha)
=\begin{pmatrix}
  0 & e^{-2i\alpha}(J_{xx}-iJ_{xy}) \\ e^{2i\alpha}(J_{xx}+iJ_{xy}) & 0
\end{pmatrix}
\label{Eq:3}
\end{equation}
where $\mathbf{\overline{J}}_{c}(\alpha)$ connects the incident circular polarized Jones vectors to the scattered circular polarized ones. When $J_{xy}=J_{yx}=0$ by mirror symmetry~\cite{PRA_Jones_Calculus}, the scatterer flips the polarization state of an input beam from left (right) to right (left) circular polarization~\cite{APL_spin_to_orbital,LSA_spin_to_orbital}. Simultaneously, an additional uniform phase factor $e^{\pm 2i\alpha}$ called Pancharatnam-Berry phase~\cite{kang2012twisted} is introduced, which is able to produce an OAM value of $\pm 2q\hbar$.

Ideally, one can obtain a perfect (100\%) conversion if $J_{xx}$ and $J_{yy}$ have the same unit amplitude and 180-degree phase difference~\cite{lei zhou}. It is well known that PEC and PMC surfaces reflect EM waves perfectly but with a reverse phase. If a metasurface functions as a PEC plane for $x$-polarized EM waves, then we got $J_{xx}=-1$. Likewise, if the metasurface functions as a PMC plane for $y$-polarized EM waves, then we arrive at $J_{yy}=1$. Therefore, a mirror-symmetric and anisotropic PEC-PMC scatterer will achieve 100\% efficiency for the OAM conversion. Inspired by this concept, we propose a novel metasurface as shown in Fig.~\ref{Fig:unit_cell}. Figure~\ref{Fig:unit_cell}(b) presents a scatterer of the metasurface comprising two dielectric layers, two artificial metal surfaces, and one ground plane. Periodic boundaries and Floquet ports are imposed respectively at the lateral and longitudinal sides of the scatterer.

The top-right inset in Fig.~\ref{Fig:unit_cell}(c) shows the artificial anisotropic PEC surface. Each metal strip with a width of $t$ is separated by a gap $g$. The metal strip array behaves like a parallel-plate waveguide. Plane waves polarized along the $y$ direction freely pass through the strip array, because there is no cutoff frequency for the excited TEM mode. While for $x$-polarized plane waves, TE modes need to be considered, which have cutoff frequencies. Here we choose a sufficiently small gap so that the operating frequency is well below the cut-off frequency of the fundamental $\text{TE}_1$ mode. By employing the artificial PEC surface, the $x$-polarized plane wave is totally reflected with an offset phase of $\pi$. The bottom-right inset in Fig.~\ref{Fig:unit_cell}(c) is the artificial PMC surface realized by the mushroom-like high-impedance surface~\cite{sievenpiper1999high}. A via inductor links the square patch to the ground plane. The gap capacitance exists between adjacent patches. When the mushroom structure is on resonance, the formed high-impedance surface can be regarded as a PMC plane. In view of a fact that the PEC surface is on top of the PMC surface, the $x$ polarized wave is perfectly reflected back by the PEC surface ($J_{xx}=-1$), and the $y$ polarized wave passing through the strip array is then perfectly reflected back by the PMC surface with a zero phase shift ($J_{yy}=1$).

\section{Results}

\subsection{Metasurface with azimuthally continuous loops}
\begin{figure}[!tbc]
\centering
\includegraphics[width=3.3in]{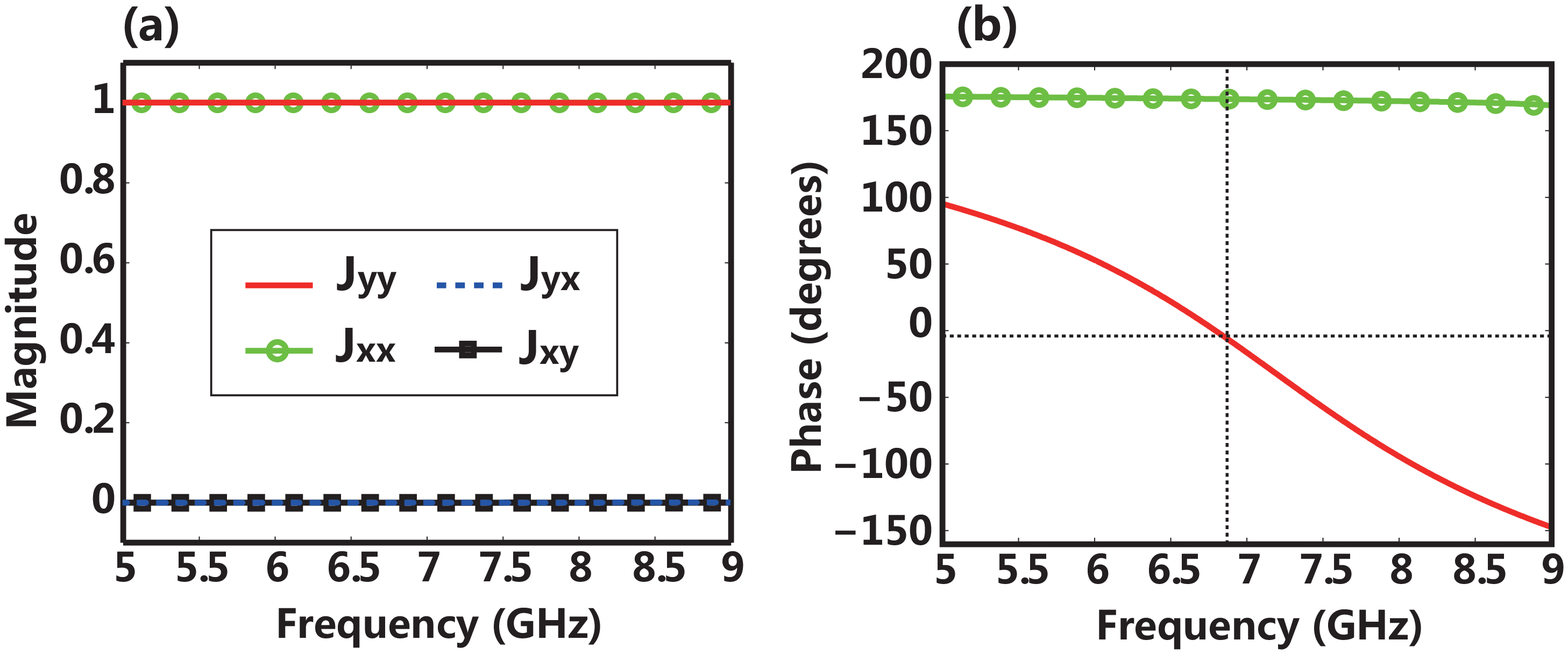}
\caption{Simulated reflection coefficients of a scatterer in the proposed metasurface at Fig.~\ref{Fig:unit_cell}. (a) magnitude; (b) phase.}
\label{Fig:unitcell_result}
\end{figure}

All the simulations are conducted by the commercial software CST MWS. Figure~\ref{Fig:unitcell_result} shows the simulated reflection coefficients of the scatterer as depicted in Fig.~\ref{Fig:unit_cell}(b). As expected, the amplitudes of the co-polarized reflection coefficients are 1, while the amplitudes of the cross-polarized reflection coefficients are zero due to the symmetric scatterer. In contrast to the strip array with a constant reflection phase of 180 degree, the mushroom structure behaves like an LC resonator with a varying reflection phase. The expected 180-degree phase difference can be realized at $6.9$ GHz.

In order to generate OAM of $\pm 2\hbar$ with a helical wavefront, the reflection phase shall be varied from 0 to 720 degrees, and the reflection amplitude is required to be a constant at different positions of the metasurface. The varying phase, as described in Eq.~\eqref{Eq:3}, is fulfilled by azimuthally rotating the scatterers (or manipulating the orientations of the scatterers). Considering the mushroom structure functions as an isotropic PMC surface, no rotation implementation is needed. For the PEC surface, if the length of the strip array along the $x$ direction is infinitely small, rotation of the tiny strip scatterers will lead to a series of concentric loops. The spacing between adjacent loops is just the strip gap $g$ as illustrated in Fig. \ref{Fig:unit_cell}(c). Interestingly, the composite PEC metasurface (concentric loops) does not have any geometric discontinuities along the azimuthal direction. The whole PEC-PMC anisotropic metasurface [See Fig. \ref{Fig:unit_cell}(a)] is rotationally invariant about the concentric origin. Due to the angular momentum conservation by rotational invariance, the metasurface allows spin-to-orbit coupling, where a change of the SAM will modify the OAM. In other words, the metasurface perfectly converts a left (right) circularly polarized plane wave carrying $\pm \hbar$ SAM to a right (left) circularly polarized vortex beam carrying $\mp \hbar$ SAM and $\pm 2\hbar$ OAM.

\begin{figure}[!tbc]
\centering
\includegraphics[width=3.0in]{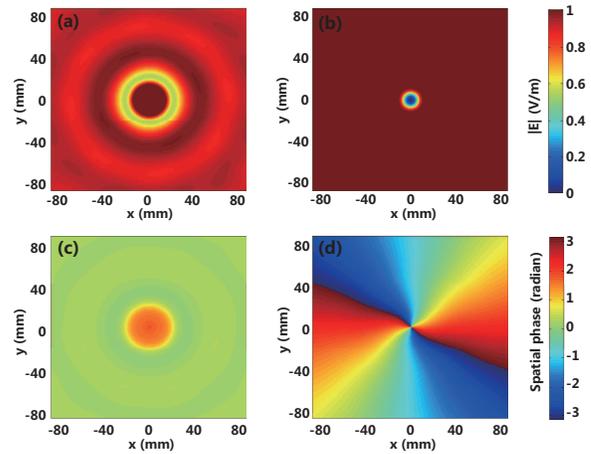}
\caption{The amplitude and phase distributions of electric fields at a transverse plane. The operating frequency is 6.2 GHz. The plane is 10 mm above an ideal metasurface. The metasurface is constructed by concentric loops on top of an ideal PMC plane. (a, b) amplitude; (c, d) phase; (a, c) incident electric fields; (b, d) reflected electric fields.}
\label{Fig:pecpmc}
\end{figure}

For simplicity and a comparative study, we first simulate concentric loops on top of an ideal PMC plane. The incident wave is a right circularly polarized plane wave at the frequency of $6.2$ GHz. Figure~\ref{Fig:pecpmc} shows the amplitude and phase distributions of electric fields at a transverse plane of $z=10$~mm. The incident and reflected fields are extracted by projecting the total electric field onto two orthogonal circular-polarized Jones vectors. A clear electromagnetic vortex (phase singularity) is observed at the origin of the reflected field as shown in Fig.~\ref{Fig:pecpmc}(b). The field amplitudes for both incident and reflected waves are around $1$~V/m, indicating that the reflection amplitude is $1$ (where the dielectric losses of substrates are ignored). For the reflected field, phase continuously increases from $0$ to $4\pi$ around the vortex. Integrating the phase of the reflected field around a path enclosing the vortex yields an integer multiple of $2\pi$. This integer $2$ is known as the topological charge of the OAM beam, which is generated by the PEC-PMC metasurface with the topological charge of $1$.

\begin{figure}[!tbc]
\centering
\includegraphics[width=3.0in]{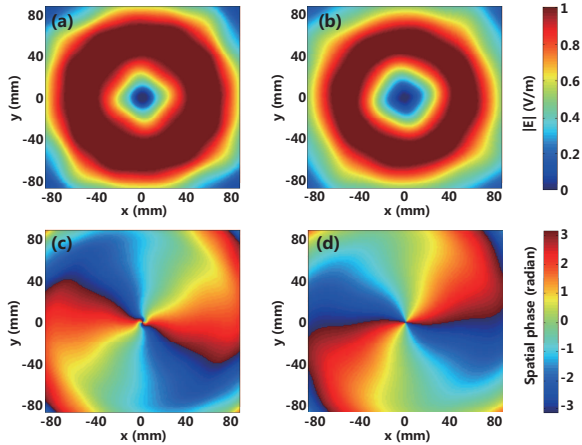}
\caption{The position-dependent amplitude and phase distributions of reflected electric fields at a transverse plane. The operating frequency is 6.2 GHz. The plane is $z$ mm above the proposed metasurface at Fig.~\ref{Fig:unit_cell}. (a, b) amplitude; (c, d) phase; (a, c) $z=10$~mm; (b, d) $z=20$~mm.}
\label{Fig:pecmushroom}
\end{figure}

Figure~\ref{Fig:pecmushroom} shows the position-dependent amplitude and phase distributions of reflected electric fields away from the proposed metasurface (See Fig.~\ref{Fig:unit_cell}). The observation planes are placed at $z=10$~mm and $z=20$~mm above the metasurface. After comparing Fig.~\ref{Fig:pecmushroom} to Fig.~\ref{Fig:pecpmc}, the maximum amplitude, electromagnetic vortex and helical phase front obtained by the proposed metasurface are almost identical to those by the ideal metasurface. This strongly confirms that the proposed design generates OAM with a nearly perfect conversion efficiency. Moreover, for the $z=10$~mm case, the abnormal phase distribution near the vortex is caused by undamped evanescent modes excited by the mushroom structure. When the distance between the observation plane and metasurface becomes large, the evanescent modes are significantly decayed. Under this situation, the reflected field from the mushroom structure is almost same as that from the ideal PMC plane.

\begin{figure}[!tbc]
\centering
\includegraphics[width=3.0in]{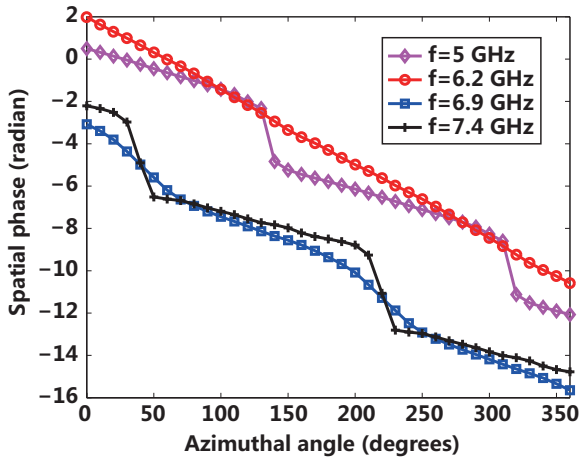}
\caption{The frequency dependence of phase distribution along the azimuthal coordinate at a constant radius of $50$ mm. The observation plane is 10~mm above the proposed metasurface at Fig.~\ref{Fig:unit_cell}.}
\label{Fig:pecmushroom_f}
\end{figure}

Figure~\ref{Fig:pecmushroom_f} illustrates frequency-dependent phase distribution along the azimuthal coordinate at a constant radius of $50$ mm. The observation plane is placed at 10~mm above the proposed metasurface. A desired linear dependence between the spatial phase and azimuthal angle is achieved at $6.2$ GHz. The operating frequency is shifted from $6.9$ GHz for the scatterer as depicted in Fig.~\ref{Fig:unitcell_result} to $6.2$ GHz for the whole PEC-PMC metasurface. Different from the scatterer configuration, in the composite metasurface, the concentric loops in the circular cylindrical space are not align to the periodic mushroom structure in the Cartesian space (See Fig.~\ref{Fig:unit_cell}). Hence, original Floquet modes in the scatterer with lateral periodic boundary conditions are broken, which is responsible for the frequency shifting. Moreover, the frequency-dependent phase-azimuthal angle diagram offers a powerful tool to examine the generated OAM pattern.

\subsection{Metasurface with azimuthally discontinuous loops}
\begin{figure}[!tbc]
\centering
\includegraphics[width=3.5in]{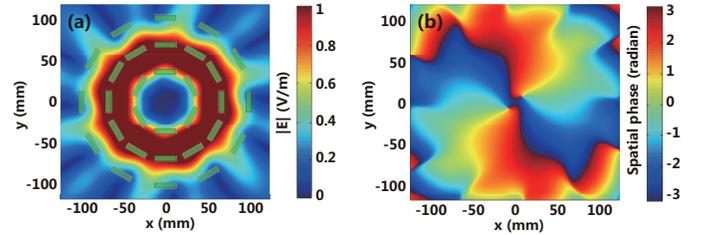}
\caption{The amplitude and phase distributions of reflected electric fields at a transverse plane $z=40$~mm. The operating frequency is 6.2 GHz. The metasurface is constructed with azimuthally discontinuous dipole scatterers (with a topological charge of $q=1$) placed above an ideal PMC plane. (a) amplitude; (b) phase.}
\label{Fig:l1}
\end{figure}

For comparison, an azimuthally discontinuous metasurface is constructed by similar PEC-PMC scatterers. The dipole PEC scatterers are depicted in Fig.~\ref{Fig:l1}(a); and the reflected field is recorded at a transverse plane of $z=40$~mm. To make a fair comparison, the distribution of the dipole scatterers satisfies the same rotational symmetry as the concentric loops (See Fig.~\ref{Fig:unit_cell}). In this simulation, we make use of an ideal PMC at the bottom to better clarify the influence of the discontinuous dipole scatterers. The operating frequency of $6.2$~GHz is identical to the concentric loops (See Fig.~\ref{Fig:pecpmc}). The field pattern shows the azimuthally discontinuous loops also obtain the vortex beam carrying the OAM of $2\hbar$. However, due to the influence of the discontinuous scatterers, ripples are observed in the phase pattern. The ripples are caused by the $z$ component of electric field generated from the edges of the discontinuous scatterers. The distortions will become more serious when the observation plane moves closer to the discontinuous metasurface. Particularly, the $E_z$ component of near-field can be completely eliminated by using the continuous loops. For the metasurface with the continuous loops, no resonant scatterers exists and thus smooth phase distribution can be achieved at both near- and far-field regions.

\begin{figure}[!tbc]
\centering
\includegraphics[width=3.5in]{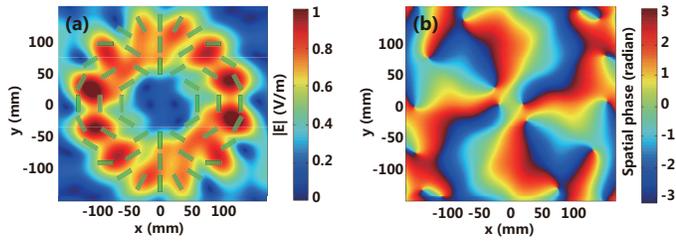}
\caption{The amplitude and phase distributions of reflected electric fields at a transverse plane $z=100$~mm. The operating frequency is 6.2 GHz. The metasurface is constructed with azimuthally discontinuous dipole scatterers (with a topological charge of $q=2$) placed above an ideal PMC plane. (a) amplitude; (b) phase.}
\label{Fig:l2}
\end{figure}

By using the discontinuous dipole scatterers, vortex beams with high-order OAM can be generated. Figure~\ref{Fig:l2}(a) shows the schematic pattern. Obviously, the topological charge of the metasurface is $2$ and that of corresponding OAM beam is $4$. In Fig.~\ref{Fig:l2}(b), the spatial phase experiences a linear increase from $0$ to $8\pi$ along a full circular path.

\section{Conclusion}
In conclusion, we propose two types of PEC-PMC anisotropic metasurfaces for generating vortex beams carrying OAM at the microwave frequency. Both types of metasurfaces achieve nearly $100\%$ conversion efficiency. One metasurface is composed of azimuthally continuous loops, which achieves smoother phase distribution than the other metasurface with azimuthally discontinuous dipole scatterers. The continuous metasurface converts a left (right) circularly polarized incident plane wave with zero OAM to a right (left) circularly polarized reflected vortex beam with $\pm 2\hbar$ OAM. Furthermore, the discontinuous metasurface generates the vortex beam with arbitrary-order OAM. Our work provides a great convenience to high-efficiency OAM generation. In future, we could explore emerging digital metamaterials~\cite{tiejun cui} to generate vortex beam with arbitrary topological charges.

\section*{acknowledgements}
This work was supported by the Research Grants Council of Hong Kong (GRF 712612 and 711511), National Natural Science Foundation of China (Nos. 61271158 and 61201122), Seed Fund of University of Hong Kong (Nos. 201309160052 and 201311159108), and University Grants Council of Hong Kong (No. AoE/P-04/08). The authors also thank the collaborator HUAWEI technologies CO. LTD. for helping the simulation.

\end{document}